\documentstyle[12pt,epsfig]{article}
\begin{document}
\title{Detectable neutrino fluxes due to enhanced cosmic ray densities in the 
Galactic Center region}
\author{Juli\'an Candia\\
{\small\it The Abdus Salam International Centre for Theoretical Physics}\\
{\small\it Strada Costiera 11, 34014 Trieste, Italy}}
\maketitle
\begin{abstract}
We examine in detail the detectability of a signal of diffuse high energy neutrinos produced 
in the Milky Way by the interaction of cosmic rays (CRs) with the interstellar medium (ISM). 
We show that highly inhomogeneous galactic CR densities arise naturally from
large scale drift effects and the inhomogeneous distribution of CR sources.  
In particular, the CR density in the Galactic Center region 
(where the ISM density is maximal) could be an order of magnitude larger 
than the local CR density. Hence, the expected diffuse flux of neutrinos from the Galactic Center region
becomes enhanced and could be detected at energies above $\sim 10^4$~GeV 
in a km$^3$-size neutrino telescope within 10 years of operation. 
The galactic anisotropy, which is the main signature of this flux,
allows to discriminate this signal from other possible 
extraterrestrial diffuse fluxes of high energy neutrinos.
\end{abstract}
\section{Introduction}
A new generation of neutrino telescopes is now opening an exciting new window on the
high energy Universe. Indeed, planned or already built large neutrino detectors 
are capable of providing us with potentially revolutionary observations 
in the fields of particle physics, cosmology, astronomy and astrophysics.
 
Among different detection techniques proposed or already used,  
a particularly important role is played by under-ice and under-water optical \v{C}erenkov detectors. 
This technique was very successfully implemented 
in the AMANDA detector \cite{ack05} at the South Pole and in 
the Lake Baikal experiment \cite{ayn03}.
The much larger, km$^3$-size detector IceCube \cite{ahr04} is currently under construction on the Antarctic ice,
while other detectors (ANTARES, NEMO and NESTOR) are also being built in the Mediterranean Sea. 
While these under-ice/water optical \v{C}erenkov detectors
are sensitive at roughly TeV-PeV neutrino energies, which is essentially the range
of interest in this work, other experiments extend the range of sensitivity up to ultra-high
neutrino energies. Indeed, the radio \v{C}erenkov detectors (such as RICE, ANITA, GLUE and SALSA) are 
sensitive roughly in the PeV-EeV energy range, while detectors making use of the Earth's atmosphere 
as a target volume (e.g. Fly's Eye, AGASA, HiRes and Auger) are sensitive at roughly EeV-ZeV energies. 

Being able to escape even from very dense astrophysical environments,
and to travel vast distances undeflected and unattenuated, neutrinos
are a precious probe to the high energy Universe, which is
complementary to both gamma rays and cosmic rays.  
Pioneer work on high energy neutrinos produced in 
various astrophysical processes was accomplished roughly three decades ago
(see e.g. Refs. \cite{ber70,ber75,ste79}). Nowadays, a primary
motivation for studying neutrinos above TeV energies is the
identification of possible individual, point-like extragalactic
sources such as Active Galactic Nuclei (AGNs) and Gamma
Ray Bursters (GRBs). It has also been proposed recently
that jets accompanying core-collapse supernovae would produce
detectable high energy neutrino bursts \cite{and05}, which would
reveal a connection between core-collapse supernovae and long duration
GRBs. Furthermore, the existence of different kinds of
neutrino sources located within the Galaxy has been suggested and studied (see
Ref. \cite{bed05} and references therein).   
However, the identification of point sources
might be difficult to achieve, since they might be too weak to produce
unambiguous directional signals in the detectors. Hence, also the
detection of the corresponding unresolved diffuse fluxes (i.e. the
integrated fluxes produced by all sources) remains an important task.
In order to achieve this goal, a sound knowledge of all other possible
diffuse neutrino signals is obviously crucial.

At low energies, it is well known that {\it conventional} neutrinos produced 
by the decays of pions and kaons due to cosmic ray (CR) interactions 
in the atmosphere give the dominating signal \cite{vol80,gai02} 
(and this is, in fact, the only type of signal identified so far). 
At energies above few hundreds of GeV,
however, this component becomes strongly attenuated due to the fact
that the parent mesons live longer and interact before
decaying. The picture then looks more confusing, since different 
high energy neutrino fluxes of diverse origin, but possibly similar intensity, 
are expected to show up \cite{lea00}.

Besides the extragalactic fluxes mentioned above, other expected diffuse components of
comparable (or even larger) intensity are, for instance, the so-called
{\it prompt} fluxes arising from the decay of short-lived charmed
particles produced by CR interactions in the atmosphere
(see e.g. Refs. \cite{thu96,mar03,can03b}). The evaluation of this
prompt component requires taking into account next-to-leading-order
processes in the charm production cross section, which is also 
dependent on the behavior of the parton distribution functions at very
low values ($x<10^{-4}$) of the fraction of momentum carried by the
partons, well below the range measured in present accelerators. The
large uncertainties coming from the small-$x$ extrapolation procedures are
propagated on the estimated prompt neutrino fluxes and span more
than an order of magnitude. 
A characteristic signature of this (roughly isotropic) signal is the flavor ratio
$\nu_e:\nu_\mu:\nu_\tau\sim 1:1:{{1}\over{10}}$, clearly different
from the ratio 1:1:1 expected in a standard scenario for any
extraterrestrial signal, due to the effects of neutrino flavor 
oscillations during propagation \cite{ath00}.

Another contribution to the high energy diffuse neutrino fluxes is that
resulting from the interactions of the CRs present all over the Galaxy
with the ambient gas in the interstellar medium (ISM)
\cite{ber75,ste79,dom93,ber93,ing96}. The very low densities in the ISM
($\rho_{ISM}\simeq 1/{\rm cm}^3$) imply that essentially all mesons
produced decay in this case without suffering any attenuation, and
hence the neutrino fluxes are just the conventional ones from $\pi$,
$K$ and $\mu$ decays.  Since these neutrino fluxes depend on the
column density of the ISM along the particular direction considered,
this component shows a significant anisotropy, being maximal in the
direction to the Galactic Center (GC) (and generally enhanced near the
Galactic Plane), while being minimal in the direction orthogonal to it.

Although it was recognized that these diffuse fluxes of
Galactic neutrinos could compete with the other expected contributions at
high energies, a careful investigation of this signal was missing in
the literature. Indeed, inhomogeneities in the CR density at different
locations in the Galaxy were disregarded in previous investigations
\cite{ste79,can03b,dom93,ber93,ing96,bea04}, in which variations in the 
ISM density were generally neglected as well. On the contrary, in this
work it is shown that large CR density inhomogeneities arise due to both 
large-scale drift effects and the inhomogeneous distribution of CR sources,  
producing an order-of-magnitude enhancement of the CR density in the GC region, 
where the ISM density is also maximal. Hence, this effect 
enhances the fluxes expected from the GC region for the diffuse signal of
neutrinos produced in the Milky Way.

A detailed study of this contribution is important on its own, since
its detection would imply the discovery of a new high energy neutrino
signal, which would be also a new probe of cosmic rays and the
interstellar medium. But, furthermore, it is also crucially important
as a background in order to identify a real extragalactic flux, and
hence a new probe of the high energy Universe.  As already pointed
out, measurements of the anisotropy in galactic coordinates could be
used to distinguish between these different possibilities.

In the next Section, we focus on the diffusion/drift scenario that
describes the transport of CRs in the Galaxy under turbulent
conditions \cite{ptu93,can02a,can02b,can03a}.  While obtaining a sound
agreement with local CR observations, we show that large
inhomogeneities in the galactic CR density do arise as well.  
In Section 3, we calculate the relevant 
neutrino fluxes and corresponding event rates, showing that the enhanced Galactic 
flux produced in the GC direction would dominate at high energies above $\sim 10^4$~GeV. 
Despite the small event rates, the detection of the diffuse Galactic signal 
in a km$^3$-size telescope such as IceCube is shown to be feasible and realistic. 
Finally, the Conclusions are presented in Section 4.

\section{The transport of cosmic rays in the Galaxy}

According to the standard picture, galactic CRs are accelerated in
supernova remnants, the expanding
shock wave fronts produced by the explosion of supernovae.
This mechanism is suitable for producing high energy CR fluxes with power-law
spectra. The different nuclear components generated in this way may have upper 
cutoffs at the energies $E_{cZ}\simeq ZE_c$, where $E_c$ would correspond to
the cutoff of the proton component. In the context of the supernova paradigm 
for the origin of galactic cosmic rays, this upper cutoff may range between 
$E_c\simeq 10^{15}$ and $10^{17}$~eV, depending on whether the strong shocks
develop in the ISM or in the stellar wind of the predecessor star \cite{bie95}. 
Higher cutoffs can be achieved, for instance, in a cooperative
acceleration scenario, in which the collective effect of multiple stellar winds gives
rise to an additional acceleration process \cite{anc04}. For the sake of simplicity, 
here we assume a featureless production of galactic CRs (ignoring cutoffs) and 
study how the spectrum is shaped by the CR turbulent transport. 
Anyhow, the neutrino fluxes at energies above the PeV are too faint to be 
detectable (see the expected event rates in Fig.5), and hence the high 
energy end of the galactic CR spectrum should not be relevant for the 
present discussion.     

In order to characterize the transport of cosmic rays in the Galaxy, we will
present here a simplified analytical realization of the 
diffusion/drift scenario \cite{ptu93,can02a,can02b,can03a}.  
It should be stressed that the previous investigations on this scenario 
were concerned only with the {\bf local} observations
of CRs, proposing suitable explanations of the CR spectrum and the
observed knees, as well as of the composition and anisotropy measurements. 
Here we focus our attention instead on the {\bf global} inhomogeneities 
of the CR population in the Galaxy arising from the inhomogeneous distribution 
of sources and from the onset of macroscopic large-scale drifts.  
Indeed, the main goal of this work is to show the relevant role played 
by these global inhomogeneities in the production of detectable neutrino fluxes 
from the GC region. 
In order to check the consistency and suitability of the approach used here, the
model results for the local CR spectrum are required to agree with the corresponding 
experimental observations.  
 
The region of propagation of the CR particles, the Galaxy, is often
assumed for simplicity as a cylinder of radius $R$ and height
$2H$. For definiteness, here we will consider a flat model of the
Galaxy with $R=20$~kpc and $H=2$~kpc.  The galactic disk, which
contains most of the visible matter, has a height of a few hundred pc
and is embedded in a much larger structure, the galactic halo,
which is thought to contain large amounts of dark matter. Associated
to both the disk and the halo there are large-scale regular magnetic
fields of a typical intensity of a few $\mu$G, which are oriented to a
good approximation in the azimuthal direction.  
In the following, a single extended regular magnetic field ${\bf B_0}$
will be considered, without making any further distinction between
disk and halo components. 

The properties of this extended regular magnetic field are not well known, 
although observations seem to favor an antisymmetric structure (i.e. a field with
opposite directions below and above the galactic plane), 
as would correspond to an A0 dynamo field configuration \cite{han02}.
It is important to mention, however, that here we are adopting a simplified picture 
of the regular magnetic field. Indeed, while
Zeeman splitting \cite{yus96} and polarization observations at sub-millimeter 
wavelengths \cite{nov00,nov03} revealed very strong toroidal magnetic fields 
in the central molecular zone, other observations also indicated the presence of nonthermal radio 
filaments \cite{yus84} and threads \cite{lan99} in the Galactic Center region, which
appear to be directed nearly perpendicular to the Galactic Plane. 
Hence, the azimuthal symmetry assumed here should be regarded as a simplifying 
assumption that captures the global characteristics of the regular magnetic field in the Galaxy. 
The macroscopic drifts, anyhow, being large scale currents that dominate the diffusion of cosmic rays
at high energies, are not expected to be much dependent on the details of the magnetic
field configurations. 

Superimposed to the regular magnetic field, a random component
associated to the turbulent interstellar plasma is known to exist,
with a maximum scale of turbulence of order $L_{max}\simeq 100$~pc.
Since this random field component has a strength comparable to the
field of the regular component, namely of the order of a $\mu$G, the
transport of cosmic rays in the Galaxy takes place under highly
turbulent conditions.  Existing observational data \cite{arm81} show
that the spectrum of inhomogeneities may be the same for the density
of the gas and for the magnetic field, and that it is close to a
Kolmogorov spectrum. Hence, the random magnetic field can be assumed
to be given by a power law spectrum with Fourier components giving
rise to a magnetic energy density d$E_r/{\rm d}k\propto k^{-5/3}$, for
$k\geq2\pi/L_{max}$.

A relativistic particle of charge $Ze$ propagating across a regular
field ${\bf B_0}$ describes a helical trajectory of radius
$r_L\sin\theta$, where $r_L$ is the Larmor radius given by
\begin{equation}
r_L={pc\over Ze|{\bf B_0}|}\simeq\left({E/Z\over 10^6\ {\rm
GeV}}\right)\ \left({\mu {\rm G}\over |{\bf B_0}|}\right)\ {\rm pc}\ ,
\end{equation}
while the pitch angle $\theta$ is the angle between the regular field
and the particle velocity.  In the presence of a superimposed random
field component, the CRs scatter off the magnetic field irregularities
with associated scales of order $r_L$, changing their pitch angle but
not their velocity. This collision mechanism, known as pitch angle
scattering, provides an effective means of isotropization of the
particle trajectories as long as the condition $r_L\leq L_{max}$ is
met, which for instance corresponds to energies up to few $\times
10^8$~GeV for protons propagating in the galactic magnetic
fields. Another process known to contribute significantly to the
diffusive transport of cosmic rays is the so-called field line random
walk, which is the braiding and mixing of the magnetic field lines
themselves \cite{jok66,jok69,for74}.

Neglecting nuclear fragmentation and energy-loss processes such as ionization losses
and adiabatic deceleration, the transport of high energy galactic cosmic rays is  
governed by diffusion. It should be remarked that adiabatic losses arising
from convection in an expanding flow is an important mechanism in other contexts, e.g.
in the escape of particles from expanding supernova remnants. However, in the absence of 
strong galactic winds, this term can be safely neglected in the transport equations
describing high energy CRs confined in the Galaxy. Thus, CR propagation takes place  
according to a diffusion equation,     
\begin{equation}
\nabla\cdot{\bf J}=Q, \ {\rm with}\ J_i=-D_{ij}\nabla_j\Phi\ ,
\label{diffeq}
\end{equation}
where $Q$ describes the distribution of sources, $\Phi$ is the CR
differential flux and ${\bf J}$ the CR macroscopic current. 

The diffusion tensor is given by
\begin{equation}
D_{ij}=\left(D_{\parallel}-D_{\perp}\right)b_ib_j+D_{\perp}\delta_{ij}+D_A\epsilon_{ijk}b_k\
,
\end{equation}
where ${\bf b}={\bf B_0}/|{\bf B_0}|$ indicates the direction of the
regular field, $\delta_{ij}$ is the Kronecker delta and
$\epsilon_{ijk}$ the fully antisymmetric Levi-Civita tensor.  The
diagonal components of the diffusion tensor are $D_\parallel$ (that
corresponds to the diffusion in the direction parallel to the regular
magnetic field) and $D_\perp$ (associated to the diffusion along the
transverse directions), which depend on short scale turbulent
fluctuations, while $D_A$ is the antisymmetric (Hall) diffusion
coefficient related to macroscopic drift effects.

Detailed numerical investigations on the behavior of the
different diffusion coefficients under highly turbulent conditions
were recently performed \cite{gia99,cas02,can04}.  Although the diffusion
orthogonal to the regular magnetic field direction is typically much
slower than the parallel one (unless the turbulence level is very
high, since parallel and perpendicular motions then become similar),
both $D_\parallel$ and $D_\perp$ have the same dependence on energy
(as long as $r_L<L_{max}$, which is the case of interest for this
work).  Indeed, it is found that $D_\parallel,D_\perp\propto E^m$,
where $m$ characterizes the spectrum of the random magnetic field
energy density, given by d$E_r/{\rm d}k\propto k^{m-2}$
\cite{cas02,can04,rou03}.  In particular, $m=1/3$ for the Kolmogorov
case.  Instead, the antisymmetric diffusion coefficient has a linear
dependence on energy, $D_A\propto E$ \cite{ptu93,rou03}.  As we will
discuss further below, this stronger energy dependence of $D_A$ plays
a crucial role in the transport of high energy galactic CRs,
since the knee ($E_{k}\simeq 3\times 10^{6}$~GeV) and the second knee 
($E_{sk}\simeq 4\times 10^{8}$~GeV) of the spectrum can be explained as
due to an enhancement in the escape of CRs while going from the transverse
diffusion dominated regime (at low energies) to the Hall diffusion
(drift) dominated regime (at high energies). 

Besides the turbulent diffusion/drift scenario, many other models have been 
proposed in order to explain the CR observations around and beyond the knee 
region. For instance, it was recently proposed that most CRs at energies above $\sim 100$~TeV 
could be originated in GRBs and that the knee feature could result from the transport of CRs 
produced by a recent nearby Galactic GRB \cite{wic04}. Many other models for the origin of the knee 
are reviewed and compared in Ref. \cite{hoe04a}. 
Anyhow, the diffusion/drift scenario is among the favored models to explain the observations
of galactic CRs \cite{hoe04a}. It is based on very natural assumptions, it reproduces well the 
spectral shape and the composition observations, while it additionally predicts an anisotropy increase 
correlated with the knee of the spectrum \cite{ant04}, which is a particular feature that allows 
to single out this scenario from all other proposals.  
Further evidence to clarify the origin of the CR knees will be available from mass group spectral
observations and from more accurate measurements of the composition and the
anisotropies, as expected for instance
in the Kascade-GRANDE experiment \cite{nav04}.

Since we are here assuming the cylindrical symmetry of the system and
the regular field to be azimuthal, the parallel diffusion coefficient
plays no role and the macroscopic current is simply given by
\begin{equation}
{\bf J}=-D_\perp\nabla\Phi+D_A{\bf b}\times\nabla\Phi\ .
\end{equation}
Here, $D_A$ is considered to be a positive-definite quantity
(corresponding to the diffusion of positively charged particles),
while any sign changes arising from possible changes in the
orientation of the regular field are contained in {\bf b}. This agrees
with Refs. \cite{can02a,can02b,can04} but differs from the notation of
Ref. \cite{ptu93}.

In terms of cylindrical coordinates, Eq.(\ref{diffeq}) is explicitly
given by
\begin{equation}
\left[-{{1}\over{r}}{{\partial}\over{\partial r}}\left(rD_{\perp}
{{\partial}\over{\partial r}}\right)- {{\partial}\over{\partial
z}}\left(D_{\perp}{{\partial}\over{\partial z}}\right)+
u_r{{\partial}\over{\partial r}}+u_z{{\partial}\over{\partial
z}}\right] \Phi=Q\ ,
\label{difeqcil}
\end{equation}
where the drift velocities are
\begin{equation}
u_r=-{{\partial(D_Ab_{\phi})}\over{\partial z}}
\label{ureq}
\end{equation}
and
\begin{equation}
u_z={{1}\over{r}}{{\partial(rD_Ab_{\phi})}\over{\partial r}}\ .
\label{uzeq}
\end{equation}
Notice that, since $b_{\phi}=\pm 1$, a change in the orientation of
the regular field introduces a singular contribution to the drift
velocities.

In order to deal with analytical solutions, we will consider here a
simplified approach in which $D_\perp$ is spatially
homogeneous, while $D_A\propto r$ \cite{ptu93}. 
Let us consider a regular field antisymmetric with
respect to the galactic plane (i.e. $b_\phi={\rm sgn}(z)$).  The drift
velocities are
\begin{equation}
u_r=-2D_A\delta(z)\ ,\ \ \ u_z={{2D_A}\over{r}}\ {\rm sgn}(z)\ ,
\label{driftvel}
\end{equation} 
i.e. the vertical drifts are directed outwards the galactic plane,
while on the plane there is a singular radial drift velocity directed
towards the Galactic Center.

Since the sources are located mainly in the Galactic Plane, their
distribution is considered to be given by $Q=2h_sq(r)\delta(z)E^{-\beta}$, where 
$q(r)$ is the radial source distribution, $h_s$ is a nominal vertical scale length
and $\beta$ the index of the source differential spectrum. As said above, we will
consider that galactic CRs are accelerated in supernova remnants
(SNRs).  Hence, the source distribution will be given by the SNR
radial profile \cite{cas96}
\begin{equation}
q(r)=\left({{r}\over{r_\odot}}\right)^{\alpha_s}\exp\left(-\beta_s\
{{r-r_\odot}\over{r_\odot}}\right) \ \ \ \ ({\rm for}\ r\geq 3\ {\rm
kpc})\ ,
\end{equation}
with $\alpha_s=1.69$, $\beta_s=3.33$ and $r_\odot=8.5$~kpc.  Regarding
the SNR population close to the GC (i.e. for
$r<3$~kpc), we will assume here a constant value (that matches the SNR
population at $r=3$~kpc), since a non negligible SNR density is likely
to exist in the GC region. This assumption is indeed consistent with 
the available observations \cite{cas98}.

As the Galaxy is assumed to be flat ($H\ll R$), one can neglect the
radial gradients as compared to the gradients in $z$ and the solution
to the diffusion equation is given by
\begin{equation}
\Phi^*(E;r,z)=E^{-\beta}\left({{h_s}\over{D_A/r}}\right)
\left({{1-e^{-w(1-|z|/H)}}\over{1-e^{-w}}}\right)
\int_1^{R/r}{\rm d}y\ q(yr)\ y^{-\left(1+2/\left(e^w-1\right)\right)}\ ,
\label{singlecomp}
\end{equation}
where $w\equiv 2HD_A/rD_\perp$ \cite{ptu93}. It should be noted that
this flux, $\Phi^*$, is a single-component solution (as if all CRs
were, say, protons). Further below we consider the total CR flux,
$\Phi$, which takes into account all relevant nuclear contributions.
Also notice that, as expected from the radial drift velocity pointing
towards the GC, the density at a given radius $r_0$
receives the contribution from all sources located at larger radii,
$r\geq r_0$, while the outwardly directed vertical drifts tend to
remove the CRs from the galactic plane.  In order to make explicit the
dependence of $w$ on the CR energy, we can conveniently rewrite it
as
\begin{equation}
w=w_0\left({{E}\over{E_k}}\right)^{2/3}\ ,
\end{equation}
where $E_k=3\times 10^6$~GeV is the energy of the knee and $w_0$ a
constant parameter of order unity. To reproduce the local CR observations, 
we will assume here the plausible value $w_0=0.85$. As shown below (see Figure 2), 
this choice allows to obtain a sound agreement with experimental observations. 

\begin{figure}[t]
\centerline{{\epsfxsize=5.truein\epsfysize=3.5truein\epsffile{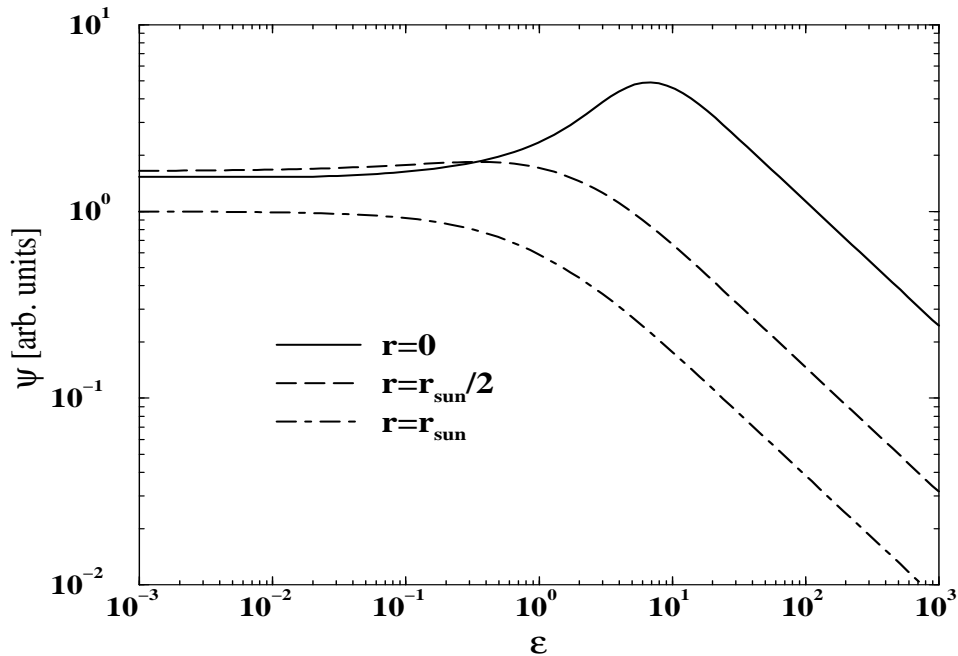}}}
\caption{Single-component CR spectra for different locations on the
galactic plane, as indicated, as functions of the adimensional,
rigidity-scaled parameter $\epsilon=E/ZE_k$ (where $E_k=3\times 10^6$~GeV).  
The normalization is taken to be unity for the low-energy local solution.
Notice the large density enhancement produced in the Galactic Center region by
the onset of macroscopic drift effects.}
\label{fig1}
\end{figure}

Figure 1 shows the single-component CR flux given by
Eq.(\ref{singlecomp}), for different locations
($r=0,r_\odot/2,r_\odot$) on the galactic plane ($z=0$), as a function
of the adimensional, rigidity-scaled parameter $\epsilon=E/ZE_k$.
Notice that $\Phi^*$ was multiplied by a factor $E^{\beta+1/3}$ in
order to produce flat solutions at low energies.  Moreover, the
normalization is taken to be unity for the low energy, local
($r=r_\odot$) solution.  For later convenience, this generic,
single-component normalized solution with zero
slope at low energies will be called $\psi(\epsilon)$.

\begin{figure}[t]
\centerline{{\epsfxsize=5.truein\epsfysize=3.5truein\epsffile{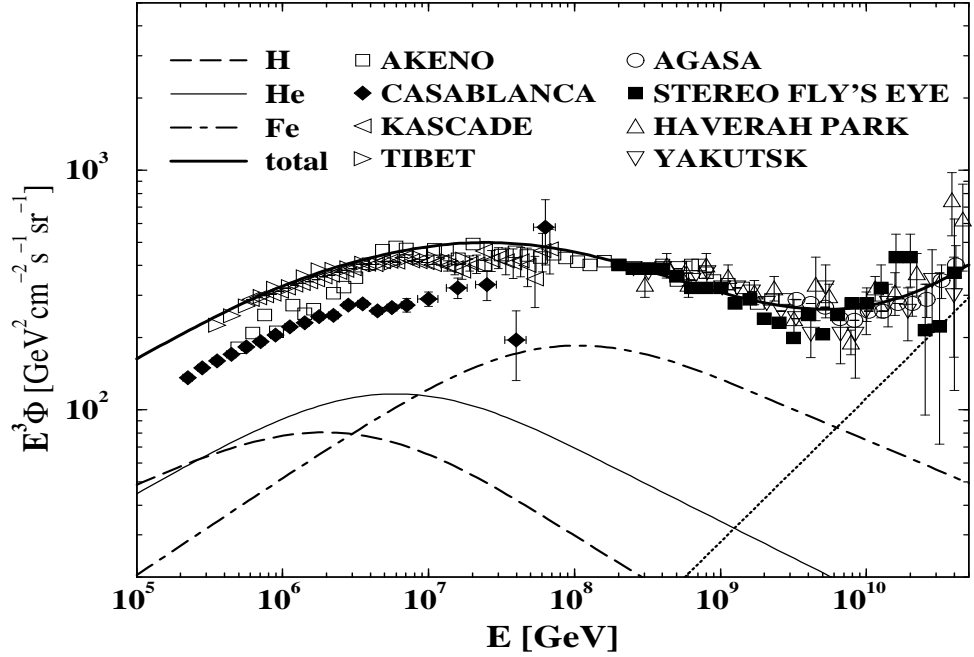}}}
\caption{Local CR spectrum in the diffusion/drift scenario
considered here, compared to experimental data sets. The main
contributions to the total flux, which correspond to nuclei of H, He
and Fe, are also indicated, as well as the assumed extragalactic flux,
given by the dotted straight line.}
\label{fig2}
\end{figure}  
\begin{table}
\begin{tabular}{|r|r|r||r|r|r||r|r|r||r|r|r|} \hline
\multicolumn{1}{|c|}{Z}&\multicolumn{1}{|c|}{$f_Z$}&\multicolumn{1}{|c||}{$\alpha_Z$}&
\multicolumn{1}{|c|}{Z}&\multicolumn{1}{|c|}{$f_Z$}&\multicolumn{1}{|c||}{$\alpha_Z$}&
\multicolumn{1}{|c|}{Z}&\multicolumn{1}{|c|}{$f_Z$}&\multicolumn{1}{|c||}{$\alpha_Z$}&
\multicolumn{1}{|c|}{Z}&\multicolumn{1}{|c|}{$f_Z$}&\multicolumn{1}{|c|}{$\alpha_Z$}\\
\hline 1 & 0.3775 & 2.71 & 8 & 0.0679 & 2.68 & 15 & 0.0012 & 2.69 & 22
& 0.0049 & 2.61 \\ 2 & 0.2469 & 2.64 & 9 & 0.0014 & 2.69 & 16 & 0.0099
& 2.55 & 23 & 0.0027 & 2.63 \\ 3 & 0.0090 & 2.54 & 10& 0.0199 & 2.64 &
17 & 0.0013 & 2.68 & 24 & 0.0059 & 2.67 \\ 4 & 0.0020 & 2.75 & 11&
0.0033 & 2.66 & 18 & 0.0036 & 2.64 & 25 & 0.0058 & 2.46 \\ 5 & 0.0039
& 2.95 & 12& 0.0346 & 2.64 & 19 & 0.0023 & 2.65 & 26 & 0.0882 & 2.59
\\ 6 & 0.0458 & 2.66 & 13& 0.0050 & 2.66 & 20 & 0.0064 & 2.70 & 27 &
0.0003 & 2.72 \\ 7 & 0.0102 & 2.72 & 14& 0.0344 & 2.75 & 21 & 0.0013 &
2.64 & 28 & 0.0043 & 2.51 \\ \hline
\end{tabular}
\caption{Cosmic ray fractional abundances (at the reference energy
$E_0=10^3$~GeV) and spectral indices from hydrogen to nickel.}
\label{Table 1}
\end{table}
 
As anticipated, the spectra show a change of slope $\Delta\alpha=2/3$
driven by the $D_A/D_\perp$ ratio. At low energies the CR transport is
dominated by the transverse diffusion, $D_\perp\propto E^{1/3}$, while
at larger energies the onset of drift effects enhances the escape as
$D_A\propto E$, hence producing a gradually steeper spectrum, which
finally changes slope by $\Delta\alpha=2/3$ in about a decade of
energy.  As will be shown further below, after summing over all CR
nuclear components in the range $1\leq Z\leq 28$ (appropriately
scaling the results according to the magnetic rigidity $E/Z$) the
envelope of the total spectrum nicely reproduces the observed knee
(which results from the escape of the light component, mainly protons
and He nuclei) as well as the second knee (which is due to the escape
of the heavy component, i.e. the Fe-group nuclei).  Another very
interesting feature of the CR spectra shown in Figure 1 is their
significant spatial inhomogeneity, especially evident at high energies
around and above the (local) knee.  Indeed, in the high energy, drift
dominated regime, the macroscopic CR currents are mainly driven by the
large scale regular magnetic fields. Hence, large accumulations (or
depletions) of CRs at different locations in the Galaxy
arise naturally by the same mechanism that is responsible for the
CR first and second knees observed locally. This phenomenon, 
being a manifestation of large-scale CR propagation properties, is 
generally also dependent on the configuration of magnetic fields 
and the distribution of sources assumed. 

In order to consider the contribution of all CR nuclear species from
hydrogen ($Z=1$) to nickel ($Z=28$), we will assume that the spectra
produced at the sources have constant (i.e. energy independent)
spectral indices, which can then be inferred from low energy direct CR
measurements. In the same vein, also the relative contribution of
different nuclear components can be inferred from low energy
observations. Hence, recalling the generic, locally normalized flux
$\psi$ defined above, the total CR flux is given by
\begin{equation}
\Phi(E)=\Phi_0\sum_Zf_Z\left({{E}\over{E_0}}\right)^
{-\alpha_Z}\psi\left(E/Z\right)\ .
\end{equation}
In this expression, $\Phi_0=3.5\times 10^{-8}\ {\rm
GeV^{-1}cm^{-2}s^{-1}sr^{-1}}$ is the total CR flux at the reference
energy $E_0$, hereafter adopted as $E_0=10^3$~GeV, $f_Z$ are the
fractional CR abundances at the same energy, and $\alpha_Z$ the low
energy measured spectral indices, which were taken from the data
compiled in Refs. \cite{wie98,hoe03} (see Table 1). Furthermore, 
we will assume an extragalactic isotropic proton component given by
\begin{equation}
\Phi_{XG}= 1.3\times 10^{-28}\left({{E}\over{10^{10}\ {\rm 
GeV}}}\right)^{-2.4} {\rm GeV^{-1}cm^{-2}s^{-1}sr^{-1}}\ ,
\label{xgflux}
\end{equation}
i.e. a flux similar to that considered in Ref. \cite{can03b}. 
Notice that an extragalactic flux, if isotropic,
permeates the Galaxy homogeneously, unless other processes different
from diffusion (e.g. reacceleration) do occur.

Figure 2 shows the total local differential CR flux, along with its
main components (protons, helium and iron), compared to data from
several experiments. The extragalactic component is also shown
separately.  As can be observed clearly, a nice agreement between
model results and experimental data is achieved even in this
simplified analytical realization of the diffusion/drift scenario, and
hence it provides a sound basis for the calculation of diffuse
galactic neutrino fluxes.  

\section{The diffuse galactic neutrino flux}  

During their confinement in the Galaxy, CRs might interact with the
ambient gas in the ISM, which is a very low density, non-relativistic
plasma constituted mainly by atomic and molecular hydrogen.  The
mesons produced in these interactions decay before losing energy in
secondary interactions, giving rise to neutrinos and photons. 
Moreover, the produced muons also decay without undergoing further energy losses
and contribute to the neutrino signal as well. 

The neutrino flux produced along a certain direction in the Galaxy is given by
\begin{equation}  
\phi_\nu(E_\nu;b,l)=\int_{E_\nu}^\infty {\rm d}E\ \sigma_{pp}^{inel}(E)\
Y_\nu(E,E_\nu)\ \int_0^{x_{max}}{\rm d}x\ \Phi_N(E,x)\ \rho_{ISM}(x)\ ,
\label{nu1}  
\end{equation}  
where $\sigma_{pp}^{inel}$ is the inelastic proton-proton cross
section, $Y_\nu$ is the neutrino yield (i.e.  the mean differential
neutrino spectrum produced in single proton-proton collisions),
$\rho_{ISM}$ the ISM nucleon number density, and $\Phi_N(E,x)$ the total 
(galactic plus extragalactic) nucleon CR flux at
the galactic location denoted by $x$. Notice that the second integral
involves the convolution of the CR flux with the ISM density along the line of
sight defined by the galactic coordinates ($b,l$).

The total proton-proton cross section in terms of the center of mass energy squared, $s$,
is parametrized by 
\begin{equation}
\sigma_{pp}\simeq\left[35.49+0.307\ln^2\left(s/28.94\
  \rm{GeV}^2\right)\right]\ {\rm mb}\ ,
\end{equation}
which combines accelerator and very high energy CR data \cite{eid04}. The inelastic cross section
is given by $\sigma_{pp}^{inel}=K\sigma_{pp}$, where $K=0.8$ is the energy independent
inelasticity used in this work. 

As regards the ISM density distribution, a model consistent with the
description given in Ref. \cite{ber93} was assumed, namely
\begin{equation}
\rho_{ISM}(r,z)=n(r)\exp(-z/0.5\ {\rm kpc})\ ,
\label{ISM}
\end{equation}
where the binned radial profiles are given in Table 2. 
It should be noticed that the ISM matter concentration peaks at the GC, 
i.e. in the same region where also the CR density is maximally enhanced.

\begin{figure}[t]
\centerline{{\epsfxsize=5.truein\epsfysize=3.5truein\epsffile{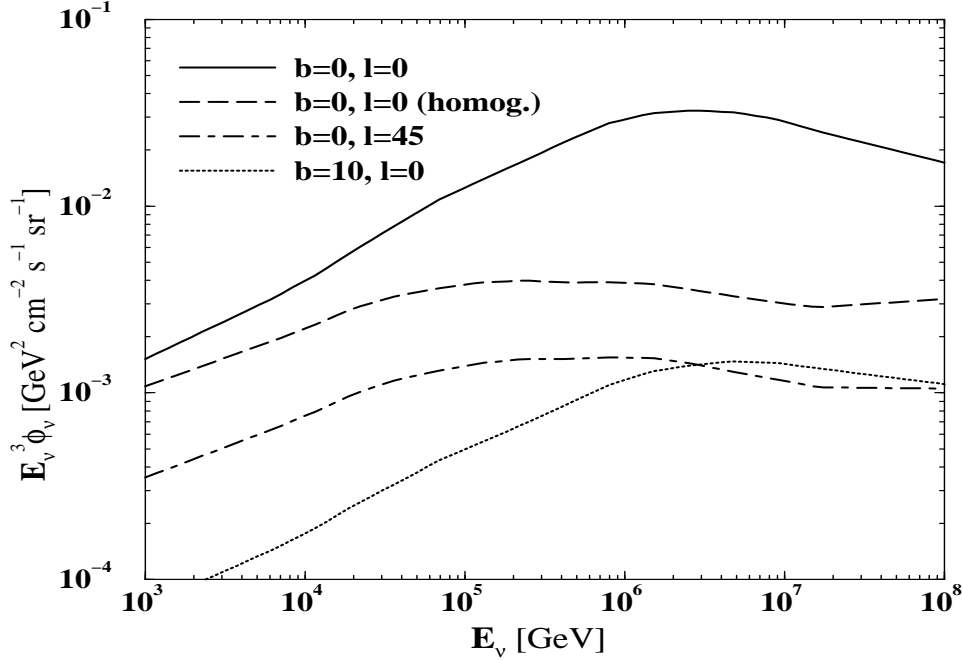}}}
\caption{Differential spectra for diffuse Galactic neutrinos arriving at the Earth from
different directions in the Milky Way. Galactic coordinates are given in degrees. 
Here and throughout, fluxes are given per flavor, but
adding neutrinos and antineutrinos. Due to neutrino oscillations during propagation,
the flavor ratio is $\nu_e:\nu_\mu:\nu_\tau=1:1:1$. For the sake of comparison, 
it is also shown the flux that would be produced from the GC direction if the CR
density were homogeneous over the whole Galaxy (i.e. just taking 
into account the local CR observations).}
\label{fig3}
\end{figure} 

\begin{table}
\begin{tabular}{|l||c|c|c|c|c|c|c|c|}\hline
\multicolumn{1}{|l||}{$r\ [{\rm kpc}]$}&
\multicolumn{1}{|c|}{$0-0.3$}& 
\multicolumn{1}{|c|}{$0.3-0.6$}& 
\multicolumn{1}{|c|}{$0.6-2$}& 
\multicolumn{1}{|c|}{$2-4$}& 
\multicolumn{1}{|c|}{$4-6$}& 
\multicolumn{1}{|c|}{$6-8$}& 
\multicolumn{1}{|c|}{$8-12$}& 
\multicolumn{1}{|c|}{$12-20$}\\
\hline {$n\ [{\rm cm}^{-3}]$}&38.1 & 2.23 & 1.91 & 1.04 & 1.14 & 0.87 & 0.40 & 0.18 \\ \hline
\end{tabular}
\caption{Binned radial distribution of the ISM density, Eq.(\ref{ISM}).}
\label{Table 2}
\end{table}

In order to calculate the neutrino yield, $Y_\nu$, we simulated fixed target 
proton-proton collisions by means of the PYTHIA 6.228 program
\cite{sjo01}, a general purpose high energy physics event generator.
This program provides an accurate representation of event properties
in a wide range of reactions, and particularly on those involving 
strong interactions, where multihadronic final states are produced. 
Lacking an exact description, the program makes use of a combination 
of analytical results and various QCD-based models, which are then implemented
to model hard subprocesses, initial- and final-state parton showers, remnants
and underlying events, fragmentation and decays, etc. 
The decay of long-lived particles, which are considered to be stable in the 
context of collider physics applications, was enabled in our simulations in order to include
the main galactic neutrino production channels. 

Figure 3 shows the differential spectra of diffuse galactic neutrinos for different
directions in the Galaxy. Here and throughout, fluxes are given per flavor, but
adding neutrinos and antineutrinos. The interactions between 
CR particles and the ISM produce roughly twice as many $\nu_\mu$ than $\nu_e$, 
but at their arrival at the Earth the produced neutrinos are evenly
distributed among the three flavors due to vacuum oscillations during propagation \cite{ath00}.
Since, as discussed above, both the ISM and the CR densities are largest in the GC region, 
the neutrino flux is maximally enhanced along the GC direction. 
Drift effects also enhance the galactic anisotropies, which are indeed 
the main signature to distinguish this diffuse galactic signal from other diffuse 
stationary extraterrestrial neutrino fluxes.
Figure 3 also shows the neutrino flux that would be produced in this 
same direction if the CR density were homogeneous over the whole Galaxy (i.e. ignoring the 
inhomogeneous distribution of CR sources and all propagation effects, and just taking 
into account the local CR observations). As anticipated, a large enhancement takes 
place at high neutrino energies above $\sim\ {\rm few}\times 10^4$~GeV. Also notice the flux increment 
taking place at very high energies for the ($b=10^\circ,l=0^\circ$) direction,
which is due to the vertical drifts that remove the CR particles from the Galactic Plane.

\begin{figure}[t]
\centerline{{\epsfxsize=5.truein\epsfysize=3.5truein\epsffile{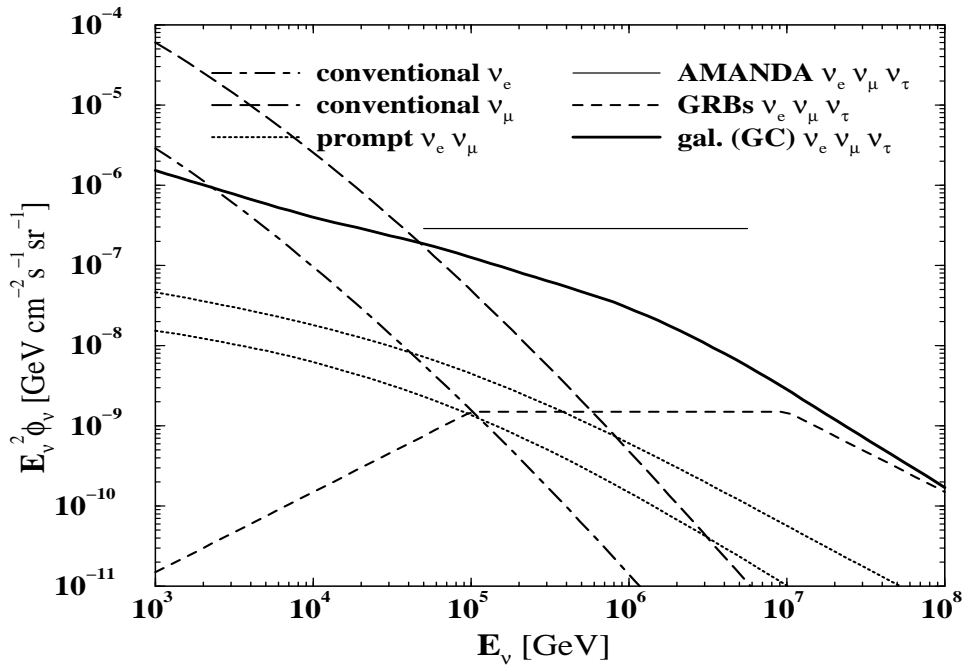}}}
\caption{Diffuse galactic signal produced in the GC direction, compared to the 
conventional atmospheric fluxes (averaged in zenith angle), the prompt atmospheric flux 
and the Waxman-Bahcall prediction for extragalactic neutrinos produced in GRBs \cite{wax99}.
Also shown is the latest AMANDA limit on the high energy neutrino flux \cite{ack05}.
The two curves for the prompt atmospheric neutrino flux indicate the adopted range 
of small-$x$ QCD uncertainties \cite{bea04}, although larger prompt fluxes are also possible.}
\label{fig4}
\end{figure} 

In order to compare the diffuse galactic signal to other high energy neutrino fluxes, 
Figure 4 shows the galactic signal produced in the GC direction, together with the 
atmospheric conventional fluxes (averaged over zenith angle), the atmospheric prompt flux 
and the Waxman-Bahcall prediction for extragalactic neutrinos produced in GRBs \cite{wax99}.
Also shown is the latest AMANDA limit on the high energy neutrino flux, obtained from their 
shower analysis \cite{ack05} and consistent with previous results from the BAIKAL experiment \cite{ayn03}. 
The two curves for the prompt atmospheric neutrino flux indicate the adopted range of small-$x$ QCD uncertainties 
(for further details, see Ref. \cite{bea04} and references therein). The prompt flux, however, could be larger 
than shown here, becoming indeed the dominating
high energy signal for directions outside the GC region \cite{bea04}.  

According to Figure 4, the galactic flux from the GC is the dominant non-conventional 
high energy neutrino signal. As shown before,
this signal is highly anisotropic in galactic coordinates (see Fig.3) and the fluxes coming from outside
the GC region are expected to fall by more than an order of magnitude. Hence, this scenario is optimistic
as regards the detection of a high energy diffuse component of extragalactic origin. In this respect, very useful 
information on the galactic diffuse background (for directions outside the GC region) can be drawn from the  
detection of the galactic signal in the GC direction.   

\begin{figure}[t]
\centerline{{\epsfxsize=5.truein\epsfysize=3.5truein\epsffile{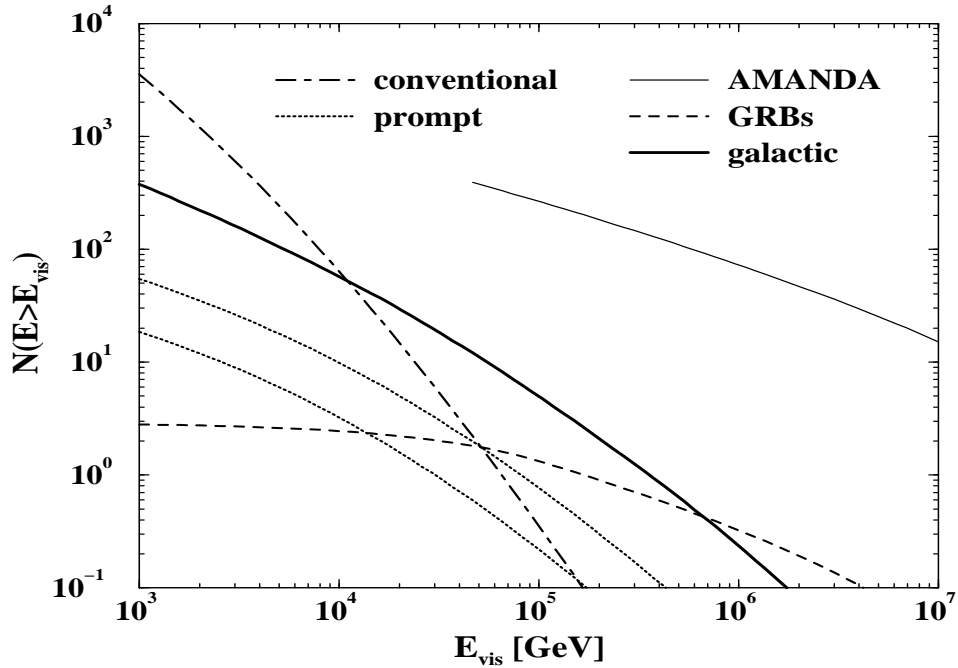}}}
\caption{Integral shower event rates as a function of visible energy $E_{vis}$, corresponding to 
the differential fluxes shown in Figure 4. The rates are calculated for a km$^3$-size detector 
taking data from the region ($|b|\leq 10^\circ,\ |l|\leq 45^\circ$) for 10 years. 
The line marked ``AMANDA" indicates the resulting integral
spectrum assuming an $E^{-2}$ power law, with no upper cutoff,
normalized by the AMANDA differential limit \cite{ack05} (which is actually given
over a slightly smaller energy range).}
\label{fig5}
\end{figure} 
Figure 5 shows the integral event rates corresponding to the fluxes given in Figure 4, as expected for a km$^3$-size detector 
observing the region ($|b|\leq 10^\circ,\ |l|\leq 45^\circ$) for 10 years of data taking. 
We focus here on shower (cascade) event rates produced by downgoing neutrinos. 
Notice that the angular range considered is consistent with the $\sim 20^\circ$ angular resolution expected 
for showers in IceCube. 
Being located at the South Pole, IceCube is capable of probing the GC region by means of
downgoing neutrinos. For TeV energies, the downgoing $\nu_\mu$ charged-current (CC) events are overcome by the 
much larger background of downgoing atmospheric muons. Instead, downgoing neutrinos can be measured by the
observation of electromagnetic/hadronic cascades produced by neutrino interactions occurring within or
near the detector fiducial volume \cite{kow03,kow05}. Furthermore, the observation of shower events 
(without lepton tracks) is particularly sensitive to the $\nu_e$ CC channel \cite{bea04} and allows  
an effective reduction of the conventional background relative to any high energy signal 
(being either the prompt atmospheric flux, or an extraterrestrial component). In fact, the visible energy 
that can be reconstructed from a shower event, $E_{vis}$, is roughly the same as the incoming neutrino energy 
for a CC $\nu_e$ interaction, while in a neutral-current (NC) interaction $E_{vis}$ is smaller by a factor 
$\langle y\rangle\simeq 0.3-0.4$. This effect, convolved with the very steeply falling incoming spectra, 
and together with the additional suppression due to the smaller NC cross sections, yields NC $\nu_\mu$ shower fluxes 
which are suppressed by roughly an order of magnitude relative to the shower fluxes arising via CC $\nu_e$
interactions. The shower detection channel was indeed recently proposed as a novel technique
to isolate the prompt atmospheric flux (see Ref. \cite{bea04}). A further analysis of the sensitivity
of the shower channel in IceCube, as well as its comparison with the complementary up- and down-going 
muon track channels, is accomplished in Ref. \cite{kow05}.
    
Despite the event rates being small, Figure 5 shows that the detection of the diffuse galactic signal 
in a km$^3$-size telescope such as IceCube is feasible and realistic. 
A background subtraction of the conventional 
flux (which can be measured with large statistics at lower energies) would allow the identification
of this relatively large galactic signal at energies above $\sim 10^4$~GeV.  

Let us finally comment on the implications of this scenario for the observations 
of galactic diffuse gamma-rays above TeV 
energies, an exciting new window on the high energy Universe opened by the measurements by the CANGAROO, 
VERITAS, H.E.S.S. and Milagro Collaborations. 
Besides the hadronic ($\pi^0$ decay) channel, the galactic diffuse photons can be produced by other 
processes such as high energy electron bremstrahlung and inverse Compton scattering with the 
interstellar radiation field.
Very recently, the Milagro Gamma Ray Observatory reported
the first observations of a diffuse Galactic Plane gamma-ray signal at TeV energies \cite{atk05}. 

Following a procedure analogous to that described at
the beginning of this Section, we calculated the photon differential flux produced by cosmic ray 
interactions in the ISM. 
In particular, the integral flux averaged over the region 
($|b|\leq 5^\circ,\ 40^\circ\leq l\leq 100^\circ$), i.e. the so-called Milagro Inner Galaxy,   
is given by $\phi_\gamma(E>1\ \rm{TeV})=1.3\times 10^{-10}$~cm$^{-2}$s$^{-1}$sr$^{-1}$, which is consistent
with the Milagro results \cite{atk05}.  
Further work on the implications of the drift-enhanced CR densities for the gamma-ray 
observations above TeV energies is under progress and will appear in a forthcoming paper. 

\section{Conclusions}

The signal of diffuse neutrinos arising from interactions between CRs and the ISM in the Milky Way was 
reconsidered here taking into account a realistic scenario for the CR transport.
The combined effect arising from an inhomogeneous distribution of CR sources (which, according to the
standard picture, are assumed as SNRs) and from the onset of CR drifts (which could be also responsible 
for the observed knees in the local CR spectrum) is shown to produce large inhomogeneities in the 
galactic CR density. In particular, this phenomenon 
could drive a large accumulation of CRs in the GC region, where the      
ISM density is also largest, thus enhancing the fluxes expected for the diffuse galactic signal of neutrinos. 
From our estimates for the event rates expected in a large, km$^3$-size detector such as IceCube,     
the detection of this signal turns out to be quite plausible.  
These prospects of detection would be even better for larger experimental facilities, 
as e.g. in the proposed IceCube Plus project \cite{hal03}.   

This neutrino flux would provide important information on the 
galactic CR puzzle and on related aspects of great astrophysical interest: the nature and the distribution of the 
CR sources, the structure and intensity of the galactic magnetic fields, the matter distribution of the ISM, etc.
Furthermore, the detection of this extraterrestrial signal could be used to test for new physics beyond the standard
model (see e.g. Refs. \cite{bea03,bea04a,bea04b}). 

The distinct signature that allows to identify this diffuse galactic flux is the large galactic anisotropy. Indeed, 
the intensity from directions outside the GC region falls below the other expected high energy
neutrino components. Hence, the detection of the galactic signal from the GC region could also 
provide very valuable information for understanding and estimating its impact as a background for measuring
extragalactic neutrino fluxes in other directions.

Furthermore, an increased population of CRs in the GC region also could have an impact on the production of neutrons
via proton-proton interactions \cite{cro05} and nuclear photodisintegration due to ambient radiation fields \cite{anc04,gra05}. 
As pointed out in Refs. \cite{anc04,cro04}, these neutron 
production and decay processes may lead to detectable gamma-ray and neutrino signals.  
Finally, let us recall that 
neutrons escaping from the GC region were also proposed as an explanation of the CR anisotropies observed 
by the AGASA and SUGAR experiments at EeV energies \cite{cro05,bos03}. However, these observations remain 
controversial, since they were not confirmed 
by the recent analysis performed with larger statistics by the Pierre Auger Collaboration \cite{let05}. 

\section*{Acknowledgments}
The author is grateful to Simona Rolli for help with the numerical calculations of this paper, 
as well as for valuable discussions. John F. Beacom, Esteban Roulet and 
Alexei Yu. Smirnov are also acknowledged for their very helpful and encouraging remarks.

\end{document}